\documentclass{PoS}

\PoS{PoS(HEP2005)171}

\usepackage{graphicx} 
\usepackage{epsfig} 
\usepackage{graphics} 

\def\beq{\begin{equation}}
\def\eeq{\end{equation}}
\def\bea{\begin{eqnarray}}
\def\eea{\end{eqnarray}}
\def\bq{\begin{quote}}
\def\eq{\end{quote}}

\def\gappeq{\mathrel{\rlap {\raise.5ex\hbox{$>$}}
{\lower.5ex\hbox{$\sim$}}}}
\def\lappeq{\mathrel{\rlap{\raise.5ex\hbox{$<$}}
{\lower.5ex\hbox{$\sim$}}}}

\def\bea{\begin{eqnarray}}   
\def\eea{\end{eqnarray}}

\title{A scenario for leptogenesis at the TeV scale}

\ShortTitle{}

\author{\speaker{Asmaa Abada}\thanks{This talk is based on work  done with M. Losada and H. Aissaou \cite{al1,al2}}\\
        laboratoire de Physique theorique, \\
        b\^at 210, Universit\'e Paris-Sud XI, \\
        91405 Orsay, France.\\
        E-mail: \email{abada@th.u-psud.fr}}

\abstract{
We propose a scenario  of thermal leptogenesis at the TeV scale in
 the context of
an extension of the Standard Model with 4 generations. We also add  one  right-handed Majorana neutrino  for each generation 
 to generate left handed neutrino masses via see-saw. The
r\^{o}le of the fourth lepton doublet is not much different from that of a
supersymmetric partner, so the leptogenesis consequences of a four generation
scenario are indicators of a more generic picture.
We obtain  a value for the baryon asymmetry of the
 Universe  in accordance with
observations by solving the full set of coupled Boltzmann equations for this
 model.}
\FullConference{International Europhysics Conference on High Energy Physics\\
		 July 21st - 27th 2005\\
		 Lisboa, Portugal}

\begin{document}

\vspace*{-1cm}We have so far two experimental evidences to call for  physics beyond the standard model (SM): the  neutrino oscillations  and the baryonic asymmetry of the universe (BAU). The SM with right-handed (RH) neutrinos ($\nu$) provides an elegant
mechanism for thermal leptogenesis.
There are two intertwined requirements: 
reproduce the spectrum for light $\nu$  in the observed range and
generate enough CP asymmetry through the out-of-equilibrium decay of heavy RH
$\nu$  (\cite{FY}). This asymmetry can be transmitted to the
baryonic sector through sphaleron induced processes to explain the BAU.  To achieve these two requirements, the RH
$\nu$ should have Majorana masses and they should couple to the
left-handed (LH) lepton doublets and the SM Higgs via complex Yukawa.
The scenario of leptogenesis occurring at the TeV scale
has been investigated for example in references 
(\cite{al1,al2},\cite{pilaftsis2}-\cite{march-russell}).  This
scale is accessible in ongoing and near future colliders. Moreover,
interesting new physics (like supersymmetry, extra dimensions, etc) could be
revealed around that scale.  It is now known that with 3 RH neutrinos it
is difficult to achieve TeV scale leptogenesis and reproduce at the same time
the small LH $\nu$ masses \cite{al1}, unless one considers quasi-degenerate
Majorana $\nu$  \cite{pilaftsis2} or small fine-tuning
\cite{Hambye2}.  There are
two approaches in the literature to realise TeV scale leptogenesis: (a)
consider a quasi-degenerate spectrum of heavy RH $\nu$ and enhance CP
asymmetry through resonant effects \cite{pilaftsis}; (b) extend the phase
space parameters, either (i) by admiting, for example, extra couplings that
allow three body decays of the RH $\nu$ leading to an enhancement of CP
asymmetry \cite{hambye3}, or, (ii) by extending the particle content. As
regards the latter possibility, one may adopt among others either of two
approaches: (1) consider a supersymmetric framework \cite{Giudice}, (2)
minimally extend the SM by having a fourth chiral generation and add a heavy
RH $\nu$ for each of the four generations, assuming that the lepton
asymmetry is due to the decay of the lightest RH $\nu$ ($N_{1}$) in the
TeV scale \cite{al1,al2}.
The scenario we present here is precisely this last approach.
It is a simple
extension of the Fukugita-Yanagida model \cite{FY} including one more generation to the SM.  We will show how this simple model which can explain the BAU and provide appropriate  values for $\nu$ masses and 
mixing angles. It is  
  based on the out-of-equilibrium decay of
a RH $\nu$, in a C and CP violating way, such that a leptonic asymmetry is produced. The latter is converted into a BAU 
due to $(B+L)$-violating sphaleron interactions which are in equilibrium above
 the electroweak breaking scale.
\vspace*{-0.4cm}\section {The scenario and its constraints}\vskip-0.4cm
The relevant part of the  Lagrangian of the model we consider is given by, 
\bea L = L_{SM} + \bar{\psi}_{R_{i}} i \partial \! \! \! /~\psi_{R_{i}} -
\frac{M_{N_{i}}}{2}(\bar{\psi}_{R_{i}}^{c} \psi_{R_{i}} + h.c.) - 
(\lambda^{\nu}_{i\alpha} \bar{L}_{\alpha} \psi_{R_{i}} \phi + h.c.) \
,\label{lag} \eea where 
$\psi_{R_{i}}$ are 2-component spinors describing the RH neutrinos 
and we define a  Majorana 4-component spinor, $N_{i} =  \psi_{R_{i}} + 
\psi_{R_{i}}^{c}$. Our index i runs from 1 to 4, and $\alpha=e,\mu,\tau,\sigma$. 
The $\sigma$ component of $L_{\alpha}$
corresponds to a LH lepton doublet which must satisfy the LEP
constraints from the Z- width on a fourth LH $\nu$ \cite{fourgen}. The 
$\lambda^{\nu}_{i\alpha}$ are Yukawa couplings and  the field $\phi$ is the
 SM Higgs boson
doublet  with vev  $v$.
We work in the basis in which the mass matrix for the RH $\nu$
$M$  is diagonal and real,
$M = diag(M_1,M_2,M_3,M_4)$ and define $m_D = \lambda^{\nu} v$.
To leading order  
the induced see-saw
$\nu$ mass matrix for the  LH $\nu$ is given by, 
$m_{\nu}  = \lambda^{\nu \dagger} M^{-1} \lambda^{\nu} v^2 $ 
which is diagonalized by the well known PMNS  matrix.

We consider the
out-of-equilibrium decay of the  lightest of the gauge singlets $N_1$.  There are two key points: (a) the CP
 asymmetry\vskip -0.2cm
\bea 
\epsilon_1 = \frac{1}{8\pi [\lambda \lambda^\dag]_{11}}
 \sum_{j\neq 1} {\mathrm Im}{ [\lambda \lambda^\dag]_{1j}^2}
 f(M_{N_{j}}^2/M_{N_{1}}^{2}),
\label{eps1}
\eea
 where $f$ is a loop factor, 
has to be magnified by the presence of a large Yukawa coupling, and (b) the
condition of out-of-equilibrium decay of ${N_{1}}$ has to be ensured, i.e.,$\Gamma_{N_{1}}={\left(\lambda^\dagger\lambda\right)_{11}}{M_1}/{8\pi}=
\sum_{\alpha=1}^{4}\frac{\left(\lambda^*_{\alpha 1}\lambda_{\alpha 1}\right)}{M_1}/{8\pi} \equiv
{\tilde {m}_{1}^{(4G)}M_1^2/(8\pi v^2)}<
 {H(T=M_{N_{1}})},
$ 
where $H$ is the Hubble expansion rate.  These conditions
restrict the size of the Yukawa couplings. \\
\\Another feature of our scenario is
that the Yukawa coupling of the heavy LH neutrino is unsupressed (order  1) to ensure that the masses of the fourth generation
leptons are heavy enough, and also to give an enhancement to the value of the 
CP-asymmetry (\ref{eps1}).
 However,  we note that the $\Delta L = 2$ processes involving the fourth family of leptons
in external legs are rapid and hence in thermal equilibrium. Consequently,
the lepton asymmetry in the fourth leptonic direction is washed out.  This 
changes the chemical equilibrium equations such that the relationship 
between  $B$ and $L$ is modified to be:
$Y_{B} = - \left({8N+4/
 (14N + 25)}\right)\sum_{\alpha=e,\mu,\tau} Y_{L_\alpha}$, 
where $Y_{L}$ is the produced leptonic asymmetry only for the light active flavours.
\vskip -1cm\section{Boltzmann Equations  and Results}
\vskip-0.4cmThe evolution of the abundance  of the $N_1$, 
 $Y_{N_1}$, and the lepton asymmetry  $Y_L$ are given by 
     \bea{dY_{N_{1}}\over dz}&=&-{z\over sH(M_1)}({Y_{N_{1}}
    \over {Y^{eq}_N}} -1)(\gamma_{_{D_1}}+\gamma_{_{S_1}}), \label{be1}\\
    {dY_L\over dz}&=&-{z\over sH(M_1)}
 \bigl[\epsilon_1 \gamma_{_{D_1}}( {Y_{N_{1}}\over Y^{eq}_N}-1)
 + \gamma_{_{W}} {Y_{L}\over Y^{eq}_L}\bigr], \label{be2}  \eea
     where $z=\frac{M_1}{T}$, $s$ is the entropy density and 
      $\gamma_{_{D_j}},\gamma_{_{S_j}}$ are the interaction 
     rates for the decay 
and $\Delta L=1$  scattering  contributions, respectively. 
 $\gamma_{_W}$ is a function of $\gamma_{_{D_j}}$ and $\gamma_{_{S_j}}$
   and 
  $\Delta L=2$ interaction rate processes, called the washout factor which is responsible 
  for the damping of the produced asymmetry. 
  In eqs. (\ref{be1}) and (\ref{be2}),  $Y^{eq}_{i}$ is the equilibrium number 
  density of
  a particle $i$.   Explicit expressions of the interaction 
   rates $\gamma_{_{D_j}}$, $\gamma_{_{S_j}}$ and $\gamma_{_W}$
    are given in details in  \cite{al2}.
An important comment is in order.  The net leptonic asymmetry is produced only in the first 3 flavours as the strong $\Delta L=2$ process involving the fourth generation washes out the asymmetry in this direction.
 
 In figure 1, we illustrate for a given set of input parameters
  the different thermally averaged reaction rates contributing to BE 
  as a function
of  $z=\frac{M_1}{T}$: $
\Gamma_{_X}={\gamma_{_X}\over n^{eq}_{N_1}}$, $
X=D,\ S, \ \Delta L=2$. It is clear from
  this plot that for this  set of parameters, and this is  true  for a wide 
  range of
  parameter space, all rates at $z=1$ fulfill the out-of-equilibrium 
  condition
  (i.e. $\Gamma_X<H(z=1)$), and so the expected washout effect due to the
  $\Delta L=2$ processes will be small.
 The generated value of the BAU is
  $\eta_B \simeq 6\times 10^{-10}$.
  Applying the see-saw mechanism to our model for the chosen values of the 
  parameters, we obtain $m_{\nu_4}>$ 48 GeV, and the three light $\nu$  masses are of the order of $10^{-1}$ eV to a few 
  $10^{-7}$ 
  eV.  
\begin{figure}
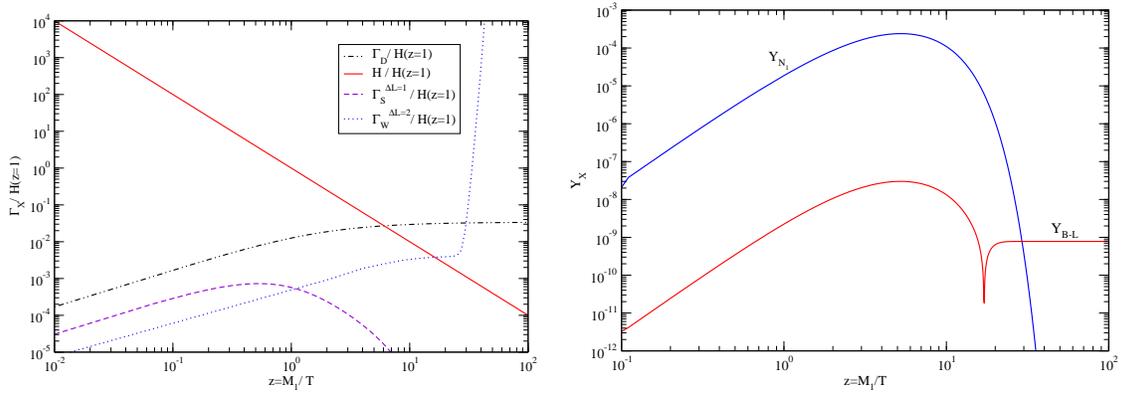
\vskip -1.5cm
\begin{center}
\begin{tabular}{ll}
 \includegraphics[scale=.3]{gamma_SDH.eps}&
\includegraphics[scale=.3]{model4G.eps}
\end{tabular}
\protect
\caption{Thermally averaged decay rates normalized to the expansion rate of the Universe $H(z=1)$ (left).  Abundance $Y_{N_1}$ and the baryon asymmetry $Y_{B-L}$ (right).}
\end{center}\vskip -1cm
\end{figure}

\section{Conclusions}\vskip -0.4cm
This scenario is an extension of the SM by
  invoking a fourth chiral family, the leptogenesis effect of which is a generic indicator of many a physics beyond the SM, e.g., 
 supersymmetry.  
We have presented the solutions to the coupled system of BE for our TeV scale model of thermal leptogenesis. We have carefully considered 
the
effect of the interactions involving the heavy fourth generation leptonic
 fields
and consistently written the BE which contribute to the final BAU together
with the appropriate conversion factor from the lepton asymmetry to the
 baryonic one.
Our results show that in this simple extension of the SM it 
is possible to produce the right amount of the BAU in a generic way.

\end{document}